\documentclass[aps,pre,onecolumn,longbibliography,nofootinbib,showkeys,showpacs,amssymb]{revtex4-1}
\usepackage{latexsym}
\usepackage{amsfonts}
\usepackage{color}
\usepackage{graphicx}
\usepackage{mathptmx}      % use Times fonts if available on your TeX system
\usepackage{bm}% bold math symbols \bm\alpha
\usepackage{enumerate}
\usepackage{subfigure}
\usepackage{listings}
\usepackage{grffile}

\begin{document}

\title{Bell's Theorem and Instantaneous Influences at a Distance}

\author{Karl Hess}
\affiliation{Center for Advanced Study, University of Illinois, Urbana, Illinois, USA}
\email{k-hess@uiuc.edu}
\thanks{Corresponding author}

\date{\today}

\begin{abstract}

An explicit model-example is presented to simulate Einstein-Podolsky-Rosen (EPR) experiments without invoking instantaneous influences at a distance. The model-example, together with the interpretation of past experiments by Kwiat and coworkers, uncovers logical inconsistencies in the application of Bell's theorem to actual EPR experiments. The inconsistencies originate from topological-combinatorial assumptions that are both necessary and sufficient to derive all Bell-type inequalities including those of Wigner-d'Espagnat and Clauser-Horne-Shimony-Holt. The model-example circumvents these inconsistencies.

\end{abstract}

\keywords{
Bell's Theorem, Instantaneous Influences, EPRB Experiments
}

\maketitle

\section{Introduction}
Einstein and Bohr debated the completeness of quantum theory and Einstein proposed a Gedanken-experiment involving two space-like separated measurement stations that demonstrated, in his opinion, that either quantum theory was incomplete or implied the involvement of instantaneous influences at a distance. This Gedanken-experiment was discussed in great detail in the literature; first by Einstein with his coworkers Podosky and Rosen (EPR) \cite{EPR}, later by Bohm and most importantly for the following considerations by Bell \cite{BI}. The Gedanken-experiment was subsequently performed in a variety of ways which are usually referred to as EPRB experiments. Three of these actually performed experiments are of particular importance: experiments of Aspect and coworkers \cite{Aspect}, because they were the first to exclude the possibility of communications between the two separated stations with the speed of light in vacuo or slower, of Weihs, Zeilinger and coworkers \cite{Zeilinger}, because their experiments were performed over the greatest distances and of Kwiat and coworkers \cite{Kwiat}, because their experiments were performed with highest statistical reliability and greatest precision.

The measurement-machinery detects a signal from a source S that is assumed to be located roughly at equal distance from both measurement stations and emanates correlated pairs of quantum entities; one part to each station. In the experiments of Zeilinger and coworkers, the space-like separation of the measurements has been as far as the separation of the islands of Tenerife and La Palma and has recently involved even satellites. The actual measurements have mostly been performed using correlated photons and the measurement equipment has involved polarizers with different measurement arrangements or ``settings" which are  denoted usually by unit vectors such as $\bf a$ at one location and $\bf b$ at the other. The following discussion, however, is not specialized to photons but admits any correlated particle-pairs and spin measurements that may also be performed by using Stern-Gerlach magnets. The same notation is used then for the orientation of these magnets. 

As mentioned, EPR \cite{EPR} intended to demonstrate that quantum mechanics is either incomplete or involves instantaneous influences at a distance. However, Bell's theorem and the violation of Bell's inequality \cite{BI} by the actual experiments appeared to deny incompleteness (the existence of so called hidden variables) and bring a decision in favor of instantaneous influences. I shall show in this paper that Bell's theorem contains two distinct claims that are based on different propositions. One of the claims is proven to contain a logical inconsistency, which becomes particularly evident from the experimental results of Kwiat and coworkers (presented in Fig.~2 of \cite{Kwiat}). As a consequence, the theorem is generally invalid and cannot decide between Einstein's alternatives. These facts are demonstrated by using an explicit model-example.

\section{ The Theorem of Bell}

{\it ``But if [a hidden variable theory] is local it will not agree with quantum mechanics, and if it agrees with quantum mechanics it will not be local. This is what the theorem says." -JOHN STEWART BELL} \cite{BT}

The word ``local" has received numerous interpretations in relation to Bell's work and his theorem may be proven in a variety of ways for different meanings of the word. It is the conviction of the author that the only acceptable meaning of ``local" clearly excludes any influences faster than the speed of light in vacuo and thus any instantaneous influences at a distance. 

A significant number of famous physicist has spoken against the possibility of instantaneous influences at a distance including Murray Gell-Mann \cite{Jaguar} and (in a much more detailed discussion) Marlan O. Scully, Yakir Aharonov and B. G. Englert \cite{Scully1}.

There have been numerous publications related to problems with Bell's inequality by notables including L.~Accardy, H.A.~De Raedt, A.~ Khrennikov,  M.~Kupczinsky, K.~Michielsen, T.~Nieuwenhuizen, W.~Philipp,  L.~Sica and others including myself. Some of the essence of their work is presented and referenced in a special issue of Open Physics \cite{Special}. \footnote{Of course, many papers have been published to praise Bell's work and to present variations on his theme, with the work of E.~Wigner being of particular importance.}

It is the purpose of this paper to show that the mathematical-physical formulation and proof of the theorem of Bell consist of two distinct parts, I and II, that are based on distinctly different premises. I shall show by example that part I of the theorem may contradict part II, because Bell's premises for the second part are much more restrictive than those of the first and include logically inconsistent topological-combinatorial assumptions. As a consequence, a way around the strictures of Bell's theorem may be found. In particular, I present an explicit model for part I that works without instantaneous influences at a distance and may be executed on two distant computer stations.

\subsection{I. Expectation value for pair measurements}

Bell's original paper \cite{BI} Introduces functions $A({\bf a}, \lambda) = \pm 1$ and $B({\bf b}, \lambda) = \pm 1$, with the important requirement that the possible outcomes $B$, symbolizing the measurements with magnet setting $\bf b$ in station 2, do not depend on the magnet setting $\bf a$ in station 1 and vice versa. Bell's $\lambda$ characterizes a pair with singlet correlations emanating from the source S. Bell regards $\lambda$ as an element of reality as defined by Mach and Einstein and also states that ``$\lambda$ stands for any number of variables and the dependences thereon of $A$ and $B$ are unrestricted." Furthermore, Bell requires that the average product of these functions over many measurements must equal the quantum mechanical expectation value for singlet spin correlations, which is: 
\begin{equation}
E({ \bf a}, {\bf b}) = -{\bf a} \cdot {\bf b}. \label{apr22n1}
\end{equation} 
Bell then states ``But it will be shown that this is not possible", a statement that forms the first part of the Bell theorem.  

These facts show that Bell committed, from the start, an inaccuracy by defining $\lambda$ as both an element of physical reality and also as a mathematical variable, in particular a variable of probability theory. The probability theory of Kolmogorov, however, is very careful to distinguish between variables and outcomes or actualizations (such as a given $\lambda_{act}$) of these variables. The actualizations are ``chosen" by Tyche, the goddess of fortune out of an ``urn" (see also \cite{HP5}). We distinguish in the following carefully between variables and possible or actual outcomes and shall use in all the derivations actual or possible outcomes corresponding to elements of reality.

As shown below, there exists no difficulty in deriving the quantum expectation value of Eq.[\ref{apr22n1}]. An additional requirement of quantum theory, however, demands that the marginal expectation values are given by $E({\bf a}) = E({\bf b}) = 0$. We show that this additional requirement can also be met in a variety of local ways.

Bell stated that it was impossible to construct a model that achieves all of these requirements of quantum theory by using functions $A$ in one wing of the experiment and $B$ in the other, which both depend solely on variables related to the local measurement stations and the emission from a common source . We refer to this statement as proposition I. 

Bell did not prove proposition I directly. He only presented illustrations of difficulties \cite{BI}  that were followed by his well known inequality, which he derived by using additional assumptions (now for the expectation values of 3 EPRB experiments). We demonstrate in a later section, by an explicit model, that these additional assumptions are generally invalid and have no bearing on the outcomes of any single-pair EPRB experiment.

\subsection{II. Three different measurement-pairs and Bell's inequality}

The second part of Bell's theorem and his proof relate to three different pairs of magnet (polarizer) settings $({ \bf a}; {\bf b})$, $({ \bf a}; {\bf c})$ and $({ \bf b}; {\bf c})$. The semicolon ``;" in between the pairs indicates that the pairs are correlated. These pairs may, in principle, be linked to the same source S or to three different sources. Kolmogorov's probability framework requires, in general, three different sample spaces and correspondingly three probability spaces; one for each magnet-setting pair. Bell, however, assumes that $\lambda$ is defined on one common probability space for all three experiments. This assumption together with Bell's particular choice of setting pairs has topological-combinatorial consequences that are necessary and sufficient (\cite{HP}, \cite{HP3}) to prove Bell's inequality:  

\begin{equation}
|E({ \bf a}, {\bf b}) - E( {\bf a}, {\bf c})| \leq 1 + E( {\bf b}, {\bf c}). \label{march25n1}
\end{equation}
We refer to this inequality together with its specific assumptions for the 3 experiments with different magnet setting pairs (explained in detail below) as Bell's proposition II.

Proposition II is in conflict with the quantum expectation values presented in proposition I and also with the results of actual EPRB experiments. Use, for example, the unit vectors ${\bf a}= [1, 0], {\bf b} = [\frac {1} {2}, \frac{\sqrt{3}} {2}]$ and ${\bf c} = [\frac {-1} {2}, \frac{\sqrt{3}} {2}]$ in Eq.~[\ref{apr22n1}] and the inequality is violated.  Variations of Bell's proof have been presented by many researchers, with the work of Wigner and d'Espagnat being of particular importance.

``Loopholes" in Bell's argument have been discussed previously by this author and coworkers. These loopholes are based on possible dependencies of Bell's functions on space-time \cite{HP} and on other globally defined variables, such as thresholds for the particle detectors \cite{DeRaedt}.  

In the present paper it is shown that proposition I is incorrect; the quantum result can be simulated by local functions as Bell required. This fact is shown by an explicit example.  Furthermore, it is proven that proposition II and Bell's inequality are marred by inconsistent topological-combinatorial assumptions and cannot be used to support proposition I, which denies the possibility to obtain the quantum theoretical result.

\section{Bell's Theorem and the Alice-Bob-tutorials}

This section is to prepare the reader for the following model of EPRB experiments and to safeguard against some common prejudices that obstruct a clear logical approach to understand such models.

\subsection{Alice, Bob and relativity}

One of the greatest obstacles for progress related to the conundrum posed by Bell's theorem is the indiscriminate use of the so called Bell game to justify Bell's theorem, a theorem of mathematical physics. Any such theorem must start from a given physical situation which is in the present case the measurement of a correlated pair of quantum particles by two separated magnets (polarizers) with directions $\bf a$ and $\bf b$ respectively. The scientist working on the theorem, be it its proof or refutation, may then use the tools of mathematics such as functions and prove from the form of the tools, e.g. the domain and the range of the functions, certain propositions. In the present case, Bell postulated that the domain of the used functions contain only variables that depend on the local physical situation in the measurement stations as well as on ``information" that is sent from a source to the stations. It is of no concern whether or not the scientists have some global knowledge while they prove or disprove a given model of mathematical physics. The ``local" quality and validity of the functions must be purely based on their mathematical form and not on what the scientists ``knew" when they developed the model.  It would be totally preposterous to call Newtons laws for the motion of the planets non-local in space and time, with the only reason that Newton knew where Mars was to be found six month later.

The well known tutorials related to Bell's theorem involve two ``players":  Alice, who has only knowledge of one wing of the EPRB experiment, and by Bob, who has only knowledge of the other wing. Reasoning involving Alice and Bob requires appropriate care. Bell himself certainly did not use any Alice-Bob arguments in his original paper which enunciated the core of his theorem. Of course, it is correct that Alice and Bob will not be able to describe the quantum correlations when they know absolutely nothing of each other. If they just measure the whole run and then combine the results by taking the count of signals on each side, a single quantum fluctuation will destroy the correlation. The true form of the correlated functions must certainly depend on the fact that they both describe the same correlated pair, but Alice and Bob can never know about the correlation without information additional to their local knowledge. Bell was not concerned about this fact when presenting his proof. He just went ahead by assuming that he was dealing with a correlated pair only and the given magnet settings in the moment of measurement. However, this assumption leads to a mathematical ambiguity in his formalism. His variable $\lambda$ and its possible outcome values are not distinctly marked as to which correlated correlated pair they belong. In the case of at least two possible magnet settings on each side, this ambiguity leads, as we will see in a later section to the logical mistake of indiscriminately pairing Bell-type functions who's $\lambda$ may actually belong to two different correlated pairs. The requirement that Alice and Bob know nothing about each other leads thus directly to the lack of the necessary mathematical distinction of variables and their possible outcomes.

In addition we need to admit the relativity of the two measurements of the EPRB correlated pairs. Consider the well known relativity-example of two elastic balls bouncing between two parallel plates each in a different inertial system; one attended by Alice, the other by Bob. Bob and Alice know nothing of each other and have the task of determining the temporal correlations between the two bouncing balls. Of course that cannot be done. If they are taking off their blindfold and are allowed to observe both balls, they may come up with a theory for the correlations that depends on the relative velocity of their moving system. Let both systems move along the $x_1$-axis of a coordinate system, one with velocity $a$ and the other with velocity $b$. The law of the correlations that Alice and Bob observe depends then on the velocity-difference $a-b$. In order to describe the law of the correlations by local functions, Alice may put ${\bf a} = 0$ and describe the movement of Bob's bouncing ball as function of any velocity $b$ of his system.  Bob may do the same with $a$ and $b$ exchanged. 

We deduce from this example that also the relativity of spin-correlation measurements in EPRB experiments does not have anything to do with distant influences but must rather be seen as a consequence of natural law. The important point is that we may indeed describe the physical events by functions of local variables from both the view of Alice and the view of Bob. The laws of physics are the same for each of them, but the physical circumstances of their respective measurements are different.  In our model below, we put the magnet setting in the left wing to ${\bf a} = [1, 0]$ and express the results in the right wing as function of arbitrary local magnet setting $\bf b$ only. Any such model must be commensurate with the relativity of all motion. In a general situation, it may thus depend even on the relative angle between the magnets (polarizers), without indicating deviations from locality. Muchowski has advanced ideas along these lines in his considerations of Bell's work as related to EPRB experiments with photon pairs~\cite{Muchowski}.  

\subsection{Completely ``random" measurements}

A particularly difficult situation for the discussions of Bell's work is created when both Alice and Bob supposedly switch their magnet settings absolutely randomly and, in addition, do not know what happens in the other wing. However, this imagined situation only obfuscates the problem and does not address the way how EPRB experiments are actually performed. The random switching done by the Aspect and Zeilinger groups does not mean that the magnet settings for the actual measurements are random. Quite the contrary,  only 3 or 4 pairs of settings are chosen in random sequence. In addition each of these sequential pairs must have at least one magnet setting in common with the other pairs. The only important randomness of the settings occurs in between the measurements, which is only relevant for reasoning about the locality of the actual experiments but not for any simulation with local functions. 

The so called ``random pairs" may thus be sorted into 3 or 4 sets, which is exactly what Wigner did in his set theoretical approach that will be discussed in a later section. It is important to realize that each of these sets concatenated by Wigner form a given sample space in the sense of Kolmogorov's set theoretic probability theory and can be, under a certain condition, regarded as a run of measurements equivalent to a completely separate EPRB experiment with a different source. The condition for the equivalence is the absence of memory effects in source and measurement equipment. Such effects are usually considered to be ``far out" and have exclusively been used to argue against Bell's inequality. 

\subsection{Counterfactuals and other issues}  

Numerous attempts have been made to prove Bell's theorem by counterfactual reasoning (that would not be permitted in the courts of law). We have shown that counterfactual reasoning does not apply when Bell's functions depend explicitly on the measurement time \cite{Counterf}. Our model presented below does exhibit such time-dependence. In addition we show under which circumstance and how counterfactual arguments may and may not be applied (see discussion of the proof of d'Espagnat).

We are not able to cover all of the issues that have been discussed in the vast literature surrounding the work of Bell and like to offer only the following observation. There are many ways  to violate Bell's inequality by admitting some global knowledge as, for example, the knowledge of the relativity of all motion. The moment, however, we exclude all global knowledge, we are only left with some magic instantaneous influences from a distantly occurring measurement.

\section{Explicit model for the quantum result}

The following explicit model may be implemented on two independent computers as well as checked by hand and represents a counterexample to Bell's claims. I refer to this model as the EQRC-model. 

In the derivation of this EQRC-model, the space-time coordinates $(x_1, x_2, x_3, x_4)$ of special relativity may be used. We use, however, for the sake of transparency exclusively coordinates of the laboratory reference frame, with $x_4$ being time-like and $x_1, x_2, x_3$ being space-like. In addition we use a global gauge field which we shall specify below. This gauge field may also be viewed as a global ``crypto key" and the whole model may be regarded as model of a computer experiment involving two (or more) computers, which have a common crypto-key that is available as a local computer-application. 
  
We assume as usual that a pair with singlet spin correlations emanates from a source in opposite directions along the $x_1$-axis. The element of reality characterizing the singlet pair is denoted by $\lambda_{st(n)}^1$, where the subscript indicates space-time coordinates $st(n)$ related to the emission from the source and $n$ is a number indicating that we deal with a correlated pair.

 The magnet (or polarizer) directions are denoted by two dimensional unit vectors $\bf a, b$ perpendicular to the $x_1$-axis and parallel to the $(x_2, x_3)$ plane. Each Stern-Gerlach magnet transmits to two detectors that are arranged perpendicular to the $x_1$-axis in the direction of $\bf a$ in the left wing (detectors $D_L^1$ and $D_L^2$) and $\bf b$ in the right wing (detectors $D_R^1$ and $D_R^2$), respectively. 

We need to decide consistently under which circumstance we regard outcomes in the two wings as equal or different (anti-correlated) in order to derive Bell-type inequalities, which can be achieved by first fixing the magnet direction (and detector alignment) of the left wing to ${\bf a} = [1, 0]$ through suitable choice of the coordinate system. Then we turn the magnet direction (and detector alignment) of the right wing such that anti-correlated outcomes are maximized (ideally occur with probability 1). 

Anti-correlated means that the detections in the left wing are registered by detector 1, while the detections in the right wing are registered by detector 2 or vice versa. We define the position $\bf b'$ that maximizes anti-correlated outcomes as the position of equal settings ${\bf b'} = {\bf a} = [1, 0]$ in the right wing. (This procedure is particularly necessary when photons, polarizers and optical fibers are involved.) We then turn the direction of the detectors in the right wing to any ${\bf b} \neq {\bf a}$ again perpendicular to the $x_1$-axis. Outcomes are defined as different, if they are registered in detectors with a different number (1, 2) or (2, 1) in the two wings. If the outcomes are with equal detector-numbers (1, 1) or (2, 2), we define the outcomes as equal. If polarizers are involved instead of magnets we need to proceed somewhat differently, but the differences matter little for the following discussions.  

All Bell-type inequalities, including that of Wigner \cite{Wigner} and d'Espagnat \cite{Espagnat}, are inequalities related to the number of equal outcomes as opposed to non-equal outcomes for Bell's three different setting pairs  (four or more pairs in the case of other inequalities).  Having in mind these details about detection, we may simplify the notation by just denoting all outcomes at detectors 1 (one in each wing) by +1 and those at detectors 2 by -1 and assign the value of +1 or -1 as the (possible) outcome for Bell's functions $A$ and $B$ respectively. If the product of the correlated outcomes in the two wings is positive, the outcomes are equal and if negative they are different. In this way both Bell's and Wigner's inequality (and all other forms of Bell-type inequalities) may be covered by the model that follows. 

Note that there is a certain arbitrariness in the definition of equal and different outcomes, because the detection of two tilt detector pairs, one in each wing, is regarded as equal if only the detections occur for the detectors with the equal number, independent of their actual setting-directions. In case one deals with more than one detector pair in each wing (as is the case for the Aspect- and Zeilinger-types of experiment), appropriate care should be taken to guarantee consistent definition of equal and different outcomes (see below).  

In addition to these conventions, we introduce a global gauge function, that is identical for all $x_1, x_2, x_3$ and varies only with the global time like coordinate  $x_4$. This function may be regarded in analogy to the concept of gauge fields in physics or, as mentioned, one may regard this function as a global crypto-key for computers, if the model is implemented as a computer experiment. The global gauge or crypto-key, both functions of space-time, or just of $x_4$ in our example, are assumed to have either no effect at all,  or alternatively, to result in a signal transfer to the alternate detector. We denote this global function by $rm(x_4) = +1$ if it has no effect and by $rm(x_4) = -1$ if it changes detectors. 

We furthermore choose for our model a very simple $\lambda_{st(n)}^1$ and let it randomly assume a value that corresponds to a real number of the unit interval, which results in $0 \leq \lambda_{st(n)}^1 \leq 1$ for each pair of measurements.

To derive the quantum result we need to consider only the outcomes for an arbitrary setting $\bf b$ perpendicular to the $x_1$-axis in the right wing. As mentioned, we choose the coordinate system of the laboratory such that  ${\bf a} =  [1,0]$. The introduction of $rm$ accomplishes vanishing marginal expectation values. 

It is not claimed that these very simplified assumptions satisfactorily simulate all aspects of natures actual mechanisms. More complicated time dependencies \cite{HP3} are certainly possible. We will see, however, that the model suffices to simulate the results of quantum theory for EPRB experiments. (To calculate the correlation with the other side quickly and explicitly from the equations given below, you may just use at first $rm = +1$.)

We assign the following possible outcomes for the functions $A$ in the left wing:
\begin{equation}
A({\bf a}, \lambda_{st(n)}^1, x_4^n) =  +rm(x_4^n) \text{ for all $\lambda_{st(n)}^1$,} \label{march26n1}
\end{equation}
where $n$ numbers the $n$'s pair of the experimental run and $x_4^n$ is the time like coordinate for the measurement of the $n$'s pair. 

For the possible outcomes $B$ in the right wing we use the arbitrary magnet setting:
\begin{equation}
{\bf b} = \frac {1} {\sqrt{(b_2^2 + b_3^2}} [b_2, b_3] .\nonumber
\end{equation}
For the possible outcomes $B$ we assign:
\begin{equation}
B({\bf b}, \lambda_{st(n)}^1, x_4^n) =  - rm(x_4^n) \label{march26n2}
\end{equation}
if we have:
\begin{equation}
\lambda_{st(n)}^1 \leq \frac {1} {2} (1 + \frac {b_2} {\sqrt{(b_2^2 + b_3^2}} ), \label{june2n1}
\end{equation}
and
\begin{equation}
B({\bf b}, \lambda_{st(n)}^1, x_4^n) =  + rm(x_4^n) \label{march26n3}
\end{equation}
otherwise. 

A little algebra shows that we obtain: 
\begin{equation}
E({\bf a}, {\bf b}) = \frac {1} {N} \sum_{n=1}^N A B = -{\bf a } \cdot {\bf b} =  - \frac {b_2} {\sqrt{(b_2^2 + b_3^2}}, \label{may6n1}
\end{equation}
the equal sign being of course only appropriate as $N$ approaches infinity. The additional requirement of quantum mechanics that the marginal expectation values $E({\bf a}) = 0$ and $E({\bf b}) = 0$ may easily be achieved by suitable choice of the function $rm(x_4^n)$. As an explicit example, one may use for $rm$ the $j$'s Rademacher function $r_j = sign[sin(2^{j+1} \pi t_n)]$, where $j = 1, 2, 3, ...$ may be chosen appropriately and $t_n$ is a dimensionless parameter corresponding to the time-like $x_4^n$.

It is important to note the following: The above formalism contains no influences from the other wing and the functions $A, B$ corresponding to Bell's functions contain only dependencies on the respective local magnet settings. The model may be generalized by replacing $rm(x_4^n)$ by the product $rm(x_4^n) rarb(x_4^n)$, where $rarb(x_4^n) = \pm 1$ is an arbitrary function of the time-like variable. Other generalizations may be used to remove asymmetries between the left and right wing. One may use, of course, a given setting in the right wing and let the left wing vary. One may even choose an infinite variety of conditioning in both wings. Such conditioning needs to be on the angle between the two magnet settings which, as we discussed in the previous section from the viewpoint of relativity, does not imply any inadmissible nonlocality.

The law of our model for the expectation values, as given by Eq.~[\ref{may6n1}], is both gauge invariant and in its form invariant to rotations of the magnet settings $\bf b$ around the $x_1$ axis. \footnote{ The quantum expectation values and corresponding quantum probability obey, of course, also a number of symmetries \cite{Huang}. It is the symmetry of the quantum probability that signals one definite distinction from general Kolmogorov probabilities.} This model, which we call the the EQRC-model refutes proposition I of Bell.

The question arises then why Bell's inequality and his proposition II appear to contradict the possibility of such a model and why proposition II is invalid. This is discussed next.

\section{Inapplicability of Bell's inequality to the EQRC-model}

In his proof of proposition II, Bell introduced three different setting pairs for magnets or polarizers. Three different equipment pairs require in general three different Kolmogorov-type sample spaces and, therefore, three different probability spaces \cite{Hbook}. Bell \cite{BI} and Wigner \cite{Wigner} (particularly in the formulation of d'Espagnat \cite{Espagnat}) created by their choice of particular setting pairs and one common probability space, unknowingly, a very restrictive and logically inconsistent topological-combinatorial situation containing a cyclicity \cite{HPA} as explained below.

\subsection{The cyclicity} 

In the notation of our EQRC-model Bell's 3 different setting pairs and possible outcomes correspond to the functions: 
\begin{eqnarray}
A({\bf a}, \lambda_{st(n)}^1, x_4^n)& \text{   ;   }& -A({\bf b}, \lambda_{st(n)}^1, x_4^n) \nonumber \\
A({\bf a}, \lambda_{st(m)}^2, x_4^m)& \text{   ;   }& -A({\bf c}, \lambda_{st(m)}^2, x_4^m) \label{april13n1} \\
A({\bf b}, \lambda_{st(k)}^3, x_4^k)& \text{   ;   }& -A({\bf c}, \lambda_{st(k)}^3, x_4^k) \nonumber 
\end{eqnarray}
where we have used the fact that $B = - A$, which was also used by Bell. If $N$ measurements are performed for each pair, we have $n = 1, 2, 3,...,N$, $m = N +1, N+2, N+3,..., 2N$ and $k = 2N+1, 2N+2, 2N+3,...,3N$. We have thus labeled the space and time related variables of the different experiments by a different number. Note that the space and time coordinates of the different experiments (symbolized by $st$ and $x_4$)  are, in general, all different.

Bell introduced now a logically and physically inconsistent assumption based on his conviction stated in his first and other papers: ``$\lambda$ stands for any number of variables and the dependences thereon of $A$ and $B$ are unrestricted." He, therefore, believed incorrectly that he needed to introduce only one symbol $\lambda$ that could stand for a whole set of variables (including space and time-like variables). Using Eq.~[12] of his original paper \cite{BI}, Bell put $\lambda$ on one single probability space and thus assumed the functions $A$ to be functions on that single probability space (see also \cite{HP3}). As a consequence and because Bell assumed that $\lambda$ could represent a set of variables, he used for each of the 3 pairs in the lines [\ref{april13n1}] the same actualization $\lambda_{act}$ of his random variable $\lambda$ and did not include any explicit time dependence. Therefore, the 3 lines [\ref{april13n1}] are reduced to:
\begin{eqnarray}
A({\bf a}, \lambda_{act}^h)& \text{   ;   }& -A({\bf b}, \lambda_{act}^h) \nonumber \\
A({\bf a}, \lambda_{act}^h)& \text{   ;   }& -A({\bf c}, \lambda_{act}^h) \label{april13n2} \\
A({\bf b},\lambda_{act}^h)& \text{   ;   }& -A({\bf c}, \lambda_{act}^h), \nonumber
\end{eqnarray}
where $h = 1, 2, 3, ..., N$. 

This procedure concatenates each 6 possible outcomes into 3, which is equivalent to assuming the existence of a joint triple probability for the possible outcomes with settings $\bf a, b, c$. The procedure fails logically, because it creates a closed loop \cite{HP3}: Two of the functions in the first two lines completely determine the functions in the third line. This is logically and physically inconsistent, because the functions of the first two lines relate to both correlated pairs in two wings and uncorrelated pairs in one wing. The third line, however, is for correlated pairs only. Bell and all of his followers have disregarded this important distinction by introducing only one common probability space. The following discussion of the experiments of Kwiat and coworkers in terms of the EQRC-model illustrates this situation clearly.

Some may still believe that Wigner's variation of Bell's inequality must hold, because it is thought to be based on set theory only. We show in the following section that it is not.

\subsection{Wigner-d'Espagnat}

Bell's derivations were investigated in great detail by Wigner and d'Espagnat, who presented a confirmation and extension of Bell's work. The derivations of Wigner and d'Espagnat seem to be based only on the rules of set theory and it is claimed in numerous publications that indeed they are.This claim, however, is false because Wigner and d'Espagnat used an assumption that lacks generality precisely as Bell's assumption does \cite{HHK2} and is also based on a mathematical mistake that I explain now using the EQRC-model.

D'Espagnat uses the 6 possible outcomes for the 3 setting pairs of Bell and transforms them into 3 triples with 9 possible outcomes. He accomplishes this transformation by adding an arbitrary third possible outcome for the magnet-setting that is not included in any of Bell's pairs. We denote these additions of possible outcomes by functions $A'$. In order to simplify the notation, we hide all variables except for the equipment settings $\bf a, b, c$.  Thus we obtain lists (columns) for the Bell-pair possible outcomes together with the added added third listings of $A'$):
\begin{equation}
A({\bf a}) A({\bf b})A'({\bf c}) \text{  ,   }A({\bf a}) A({\bf c})A'({\bf b})\text{   ,  } A({\bf b}) A({\bf c})A'({\bf a}) \label{march28n1}
\end{equation}
D'Espagnat \cite{Espagnat} and Wigner \cite{Wigner} {\it incorrectly} deduce from the existence of these sets of triples the existence of a common joint triple probability measure for all three triples of line [\ref{march28n1}]. The existence of a common joint triple probability measure and use of Bell's cyclical arrangement of settings lead immediately to the Wigner-d'Espagnat inequality,  which corresponds roughly to Bell's inequality. \footnote{ The Wigner-d'Espagnat inequality is an inequality involving the frequency of  equal and different pair outcomes for Bell's three setting pairs}

D'Espagnat's procedure, published in Scientific American \cite{Espagnat}, shows clearly how the existence of a common joint triple probability was incorrectly deduced. It is actually indeed guaranteed that a joint triple probability measure may be deduced for each separate triple of [\ref{march28n1}]. The mere fact that the triples of Eq.[\ref{march28n1}] can be listed by using the possible pair-measurement outcomes with arbitrary additions $A'$ gives us that guarantee. The mistake of d'Espagnat and Wigner was, however, that they assumed the existence of one common triple probability measure for all three triples, while in fact each of the three triples may have its own different triple probability measure. 

One can easily prove this latter fact and demonstrate the mistake by using the possible outcomes of our EQRC-model given by Eqs.[\ref{march26n2}] and [\ref{march26n3}] (with $B = -A$) and the settings ${\bf a}= [1, 0], {\bf b} = [\frac {1} {2}, \frac{\sqrt{3}} {2}]$, ${\bf c} = [\frac {-1} {2}, \frac{\sqrt{3}} {2}]$. Consider only the first two triples, use for the moment $rm = +1$ and generate the possible pair outcomes of the EQRC-model.  Second add third columns $A'({\bf c}) = +1$ and $A'({\bf b}) = +1$, respectively. Then re-install the functions $rm$ using the same $rm$ that was used for the correlated pair in each of the two respective triples. This procedure reinstates the necessary randomness everywhere. Inspection shows that the probability for all positive triple-outcomes is given by $P_{\bf abc'}(+1, +1, +1) = \frac {3} {8}$ for the first triple, while for the second triple we have $P_{\bf ab'c}(+1, +1, +1) = \frac {1} {8}$. Therefore, d'Espagnats and Wigners assumption of one common joint triple probability for the line [\ref{march28n1}] is incorrect. Many other examples may be given.

The derivations of the Bell as well as other inequalities in all textbooks are based on errors similar to that of d'Espagnat and Wigner (see for example \cite{Scully} or Norsen's table \cite{Norsen}).

  The same arguments as outlined above apply also to all other Bell-type inequalities, because they are based on joint probabilities that do not exist and are thus not applicable to actual EPRB experiments, nor to the EQRC-model.

\section{Comparison with experiments}

\subsection{Single setting in one wing}

The EQRC-model may directly be applied to simulate the very precise experiments presented in Fig.~2 of Kwiat and coworkers \cite{Kwiat}, because they chose for these results a single given setting in one wing and performed many measurements with many different settings in the other. The precision of the EQRC-model depends only on the number $N$ of simulations and agreement with the quantum result can be made as perfect as desired. Kwiat and coworkers did not include random changes of the polarizers before the registration of the actual measurement. However, there is little doubt that they would have obtained the same results if they had changed the polarizer setting just before turning to the measurement-setting. 

We may obtain from the EQRC simulations, as well as from the actual experiments, the quantum result for the setting-pairs $({\bf a}; {\bf b})$ and $({\bf a}; {\bf c})$. The averages over the outcomes with both settings $\bf b, c$ in the right wing, on the other hand, do not result in the quantum expectation value, because they do not correspond to simulations involving correlated pairs. This way Bell's inequality is naturally fulfilled, by both the EQRC-model and the actual experiments, for the ``triangles" of outcomes that exhibit equal sign for the setting $\bf a$ in the left wing and arbitrary sign for $\bf b, c$ in the right wing. There exists no contradiction here, because measurements with both the $\bf b$ and $\bf c$ settings in the right wing do not correspond to a correlated pair but to elements of reality originating from different pairs. Both the actual experiments (of Fig.~2 in \cite{Kwiat}) and the EQRC-model clearly distinguish the actual or possible outcomes corresponding to correlated and uncorrelated pairs.

However, Bell's mathematical model does not and cannot make that distinction. Bell's use of one single probability space enforces the use of identical functions for the settings $\bf b$ and $\bf c$ independent of the question of the origins of the given $\lambda_{act}^h$. Bell's identical notation for $\lambda_{act}^h$ independent of its origins from one or two different pairs represents, as far as the experiments of Kwiat and coworkers in their Fig.~2 are concerned, only a mathematical sloppiness. The application of Bell's functions to other experiments of Kwiat and coworkers \cite{Kwiat} (not presented in Fig.~2) and the experiments of the Aspect and Zeilinger groups, however, represents a serious mathematical inconsistency, because now the same mathematical abstractions $\lambda_{act}^h$ and the same functions are used for both correlated and uncorrelated elements of reality.

\subsection{Multiple settings in both wings}

Thus, experiments involving multiple settings in both wings must not be modeled using Bell's original cyclic functions, because this procedure mixes indiscriminately correlated and uncorrelated pairs. It is, of course, possible to use additional indexing and time dependencies of the functions (as done in lines [\ref{april13n1}]) to avoid the inappropriate use of a single probability space, but then it may become cumbersome to directly show the locality of the functions. The easiest way around the problems and the way that shows the locality of the procedure most directly, is probably the following: We use a symmetry law that applies to the actual experiments, the symmetry with respect to rotations around the $x_1$-axis. We, therefore consider idealized-actual experiments by rotating the actual original magnet setting such that all the left-wing magnet settings of the idealized experiments point in the $[1, 0]$ direction of a chosen coordinate system. This idealized experiment must exhibit the same correlations as the original experiment because of the existing symmetry. The technical problem with the single probability space, however, has been avoided for the idealized experiment, because we have now all different sample spaces and just have Wigner sets with a consistent notation.

{\it The magnet settings of both the actual and idealized experiments may, of course, be arbitrarily switched just before the measurement and then brought into measurement position just as the correlated pair is being registered. This whole random switching is only necessary to exclude information exchange between the two wings of the actual experiment. The functions we use in the simulations do not depend on the other wing anyway.}

Assume then that three actual EPRB experiments have the respective magnet setting pairs $([1, 0]; [\frac {1} {2}, \frac{\sqrt{3}} {2}])$, $([1, 0]; [\frac {-1} {2}, \frac{\sqrt{3}} {2}])$ and $([\frac {1} {2}, \frac{\sqrt{3}} {2}]; [\frac {-1} {2}, \frac{\sqrt{3}} {2}])$. Each pair represents one of the well known Bell-angles. We rotate then the third pair of magnets to the position of our idealized experiment $({\bf a}; {\bf c'}) = ([1, 0]; [ \frac {1} {2}, \frac{\sqrt{3}} {2}])$ and are now able to simulate the experimental outcomes with the EQRC-model and to obtain the quantum results with arbitrary accuracy for all three setting pairs.

The Aspect \cite{Aspect} and Zeilinger \cite{Zeilinger} experiments do not use Bell's 3 setting pairs but 4 settings pairs corresponding to 4 experiments with $({ \bf a}; {\bf b})$ in experiment 1, $({ \bf a}; {\bf c})$ in experiment 2, $({ \bf d}; {\bf b})$ in experiment 3 and $({ \bf d}; {\bf c})$ in experiment 4. These 4 pairs of magnet- (polarizer-)settings are used to investigate the Clauser-Horne-Shimony-Holt (CHSH) \cite{CHSH} inequality for the expectation values:
\begin{equation}
| E({ \bf a}, {\bf b}) + E({ \bf a}, {\bf c}) +E({ \bf d}, {\bf b}) - E({ \bf d}, {\bf c})| \leq 2. \label{may25n1}
\end{equation}

As above, we transform the actual CHSH type of experiments and their actual magnet (polarizer) settings into our idealized experiment by magnet rotation, now obtaining the 4 setting pairs: $([1, 0]; [\frac {1} {\sqrt{2}}, \frac {1} {\sqrt{2}}])$, $([1, 0]); ([\frac {1} {\sqrt{2}}, \frac {-1} {\sqrt{2}}])$, $([1, 0]; [\frac {1} {\sqrt{2}}, \frac {-1} {\sqrt{2}}])$ and $([1, 0]; [\frac {-1} {\sqrt{2}}, \frac {-1} {\sqrt{2}}])$. Again the EQRC-model may be applied to these setting pairs and gives the correlations of quantum theory, which violate the CHSH inequality.

Thus, according to Bell and CHSH, it is impossible to model actual EPRB experiments with local functions $A$ for certain magnet setting-pair combinations, while it is indeed possible to perform such modeling for the idealized experiments which involve the same magnet (polarizer) angles and must have the same correlations for reasons of symmetry. In fact, all of the experiments of Aspect's, Zeilinger's and Kwiat's groups show only dependencies on the angles between the two polarizers of any given experiment. The use of optical fibers in some of their experiments makes any designation of an ``absolute" angle or direction in either wing illusory.  

{\it It is instructive to imagine that Bell would have first found the EQRC-model and accepted the possibility of being able to use all local functions (particularly when accepting the relativity of all motion). He may then also have used multiple magnet-settings to produce the quantum result of the idealized experiments. Had he then rotated the magnets to turn $\bf c'$ into $\bf c$ and performed his proof for the inequality, he would have found the logical contradiction and would have been been forced to dismiss his inconsistent use of functions.}

\section{The Bell Game again and Conclusion}

Many researchers have been aware of the publications that have pointed to serious problems with Bell-type inequalities. Several of them have admitted to this author that there may be formal problems with the one or other Bell-type proof. Their deepest convictions, however, arose from the fact that no one could play the so called Bell-game with Alice and Bob \cite{Hbook}.

The Bell game and its demands highlight the crux of the epistemological questions that are going hand in hand with EPRB experiments. Some of the features of the Bell game have been described above and I add here only a few comments related to the EQRC-model. 

Alice and Bob have no knowledge of each other, particularly none of the measurement settings of the other wing and they are required to develop a theory about the possible outcomes of their local measurements. That theory needs to cover the correlations of at least the 3 different experiments with Bell's 3 setting pairs. All Alice is permitted to know are the functions $A({\bf a}, \lambda_{st(n)}^1, x_4^n) $ and her randomly chosen settings $\bf a, b$ as well as  the actualizations $\lambda_{st(n)}^1$ and $x_4^n$ etc., but she may not know of $B({\bf b}, \lambda_{st(n)}^1, x_4^n)$ and Bob's randomly chosen settings $\bf b$ or $\bf c$. The same is true for Bob, with $\bf a, b$ and $\bf b, c$ exchanged. Nor do Alice and Bob know how the pairing is actually accomplished and how the same index $n$ of a pair is actually obtained. In simple words, neither do Alice or Bob know the macroscopic machinery that deciphers the signals on the other side, nor do they know the global gauge (or global crypto-key on the computers). Of course that game cannot be played, that theory cannot be conceived. 

Some followers of Bell, however, reason that mother nature can play the game. Just let the actual measurements happen and ``put the correlated pairs together" and you will obtain the correct correlations. But how can mother nature ``know" which pairs are correlated? As mentioned, a single quantum fluctuation could falsify the pair sequence. The pairs of measurement that belong together need to be identified by some globally used space-time system and by additional measurements or assumptions that let us determine the connection between the measurement-outcomes in the space-like separated systems. How else can Alice correlate her measurements to the measurements of Bob? They both need to agree on a space-time (or space and time) coordinate system of physics that lets them determine the occurrence of the measurements and their belonging together in the different wings. This determination requires some process to identify the pairs of quantum particles and the corresponding macroscopic measurement outcomes. We have discussed this problem in a recent publication \cite{DeRaedt} and have given examples how the Bell game can indeed be played by making use of additional knowledge obtained from the particle identification method. We have conjectured that sufficient knowledge of particle and pair identification will always open a window to play the game. 

The EQRC-model uses only local functions but also does imply some global knowledge, for example the relativity of all motion. It is also compared only to idealized experiments that are constructed from the actual by applying a global symmetry law. This procedure is necessary to avoid the logical mistake inherent in applications of Bell-Wigner-CHSH-type functions, sets and inequalities.

 The game in its originally presented form just cannot be played and mother nature does not and cannot play it either. We have in this ``shaky game" \cite{Fine} the choice to admit some acceptable global information, some relative ``positioning" in a global space-time system, or to be left only with instantaneous influences at a distance as an explanation for how nature works. Such explanations are, in this authors opinion, the very last resort, because they abandon scientific method as Einstein so clearly stated by using the word ``spooky".

\end{document}